\documentclass[10pt]{article}

\usepackage[margin=1in]{geometry}
\usepackage{amsmath,amssymb,amsthm}
\usepackage{mathtools}
\usepackage{bm}
\usepackage{graphicx}
\usepackage{float}
\usepackage{caption}
\usepackage{subcaption}
\usepackage{booktabs}
\usepackage{array}
\usepackage{algorithm}
\usepackage{algpseudocode}
\usepackage{hyperref}
\usepackage{cite}
\usepackage{xcolor}
\usepackage{enumitem}

\usepackage{comment}

\newtheorem{problem}{Problem}
\newtheorem{assumption}{Assumption}

\newcommand{\R}{\mathbb{R}}
\newcommand{\Lf}{L_f}
\newcommand{\Cf}{C_f}
\newcommand{\Rf}{R_f}
\newcommand{\wstar}{\omega^*}
\newcommand{\Rload}{R_{\mathrm{load}}}
\newcommand{\vd}{v_d}
\newcommand{\vq}{v_q}
\newcommand{\id}{i_d}
\newcommand{\iq}{i_q}
\newcommand{\ed}{e_d}
\newcommand{\eq}{e_q}
\newcommand{\vdref}{v_d^*}
\newcommand{\vqref}{v_q^*}
\newcommand{\bx}{\bm{x}}
\newcommand{\bu}{\bm{u}}
\newcommand{\by}{\bm{y}}
\newcommand{\buv}{\bm{u}_v}
\newcommand{\curr}{\bm{\iota}}
\newcommand{\volt}{\bm{\upsilon}}

\hypersetup{
  colorlinks = true,
  linkcolor  = blue,
  citecolor  = blue,
  urlcolor   = blue
}

\begin{document}

\title{\textbf{Feedback Linearization and Control of a\\
Grid-Forming Power Converter in an\\
Islanded Microgrid}}

\author{Rene Ebunle Akupan$^a$, May-Win Thein$^b$ and Se Young Yoon$^a$\\
\small $^a$Department of Electrical and Computer Engineering, University of New Hampshire, Durham, NH, 03824, USA\\
\small $^b$Department of Mechanical Engineering, University of New Hampshire, Durham, NH, 03824, USA\\
\small Contact: ebrene2021@gmail.com}

\date{}
\maketitle

\begin{abstract}
In an islanded setting, grid-forming inverters must regulate their terminal voltage without support from an external grid, even though the load current depends directly on that voltage.  The usual approach is a cascaded proportional--integral (PI) controller, built on a fast inner current loop and a slower outer voltage loop, with feedforward terms used to compensate dq rotational coupling.  However, this compensation is only exact at the operating point where the controller is tuned.  This tutorial presents an alternative based on full-state feedback linearization.  It is shown that the islanded inverter model has full relative degree, which allows exact state-space linearization with
no internal or zero dynamics.  A single feedback law cancels the main nonlinear effects; rotational coupling, resistive drops, and load conductance, so that the closed-loop system behaves like two independent double integrators.  A standard pole-placement design is then used to shape the response.  The controller is tested in MATLAB against a cascaded PI baseline under identical conditions at a 20~MW operating point, including reference tracking, load step disturbances, and parameter mismatch.  The feedback-linearizing controller settles a reference step in 0.76~ms, while the PI controller does not reach the 2\% band within 50~ms.  The cascaded PI controller shows better robustness to filter parameter mismatch due to its inner-loop integral action, which reduces steady-state errors under modeling uncertainty.  Overall, the performance improvement and the robustness trade-off both come directly from the controller structures, rather than from tuning choices.
\end{abstract}


\section{Introduction}
\label{sec:intro}

The electrical power grid is undergoing a major transformation. Traditional synchronous generators are steadily being replaced by inverter-based renewable sources such as solar photovoltaic systems and wind turbines~\cite{IRENA2024}.  This shift brings new control challenges, especially in islanded microgrids, where a group of energy resources must operate independently without being connected to a larger grid.

In these islanded conditions, the inverter has to take on a much bigger role.  It is no longer just following a reference; it has to create the voltage and frequency at its terminals.  In other words, it acts as the source that everything else depends on~\cite{Mohammed2023}.  The core control problem is therefore voltage regulation, where the inverter must maintain a desired voltage magnitude and frequency, even though the load currents it supplies depend on the very voltage it produces.

The standard way to handle this problem is with a cascaded proportional-integral (PI) controller.  In this setup, a fast inner loop controls the filter inductor current, while a slower outer loop adjusts that current based on the voltage error~\cite{Bevrani2014, Zhang2021}.  The two loops are designed one after the other: first the inner loop is tuned to be fast, and then the outer loop is made slower so that the inner loop appears instantaneous.  The approach in~\cite{Guzman2025} is a typical example of this design philosophy.  In practice, the PI gains are tuned using a small-signal model, and the cross-coupling terms introduced by the dq transformation are handled through feedforward compensation rather than being built directly into the control law.

While this approach is widely used, it has some important limitations. First, the separation between the fast inner loop and the slower outer loop comes at a cost: it limits how fast the overall system can respond, since the outer loop must remain well below the bandwidth of the inner loop.  Second, the feedforward compensation used to handle cross-coupling
is only exact at the operating point where the controller is tuned, which can reduce robustness when conditions change.  Another key point is that the cascaded PI structure does not fully exploit the mathematical properties of the inverter model.  The system is input-affine and, importantly, it admits exact linearization.  This means that instead of
trying to suppress nonlinear effects with high gains and careful tuning, we can cancel them directly through the control law itself.  Doing so removes both the need for bandwidth separation and the limitations of operating-point-dependent compensation.

In this tutorial paper, such a controller is developed.  It is shown that the full-state islanded inverter model has a relative degree of two in each capacitor voltage output, with the total relative degree equal to the system order. This means full-state linearization is possible.  Using a single feedback law, the nonlinear system is transformed into two independent double integrators.  As a result, the closed-loop system has no internal dynamics, no zero dynamics, no need for an additional
outer regulator, and no requirement for bandwidth separation.  A standard pole-placement design is then used to achieve the desired performance.

This is demonstrated in MATLAB and compared with the cascaded PI baseline from~\cite{Guzman2025} under three scenarios: reference tracking, load-step disturbances, and parameter perturbations. This comparison is set up to highlight the structural advantages of the feedback-linearizing controller, rather than differences due to tuning between the controllers.

\medskip
\noindent\textbf{Notations.}
$\R^n$ denotes $n$-dimensional Euclidean space and $\R^{m\times n}$ the space of real $m\times n$ matrices.  Boldface lowercase letters denote vectors ($\bm{x},\bm{u}$) and boldface uppercase letters denote
matrices ($\bm{A},\bm{G}$).  Scalars are set in normal italic ($x,u$). The notation $\bm{P}>0$ ($\bm{P}\geq 0$) indicates a real symmetric positive definite (semidefinite) matrix.  $\bm{I}_n$ and $\bm{0}$ denote the $n\times n$ identity and zero matrices of appropriate order. $\bm{L_f} h(\bm{x})$ and $\bm{L_g} h(\bm{x})$ denote the Lie derivatives of a scalar function $h$ along the vector fields $\bm{f}$ and $\bm{g}$, respectively.

\section{System Model}
\label{sec:model}

\begin{figure}[H]
  \centering 
  \includegraphics[width=\textwidth]{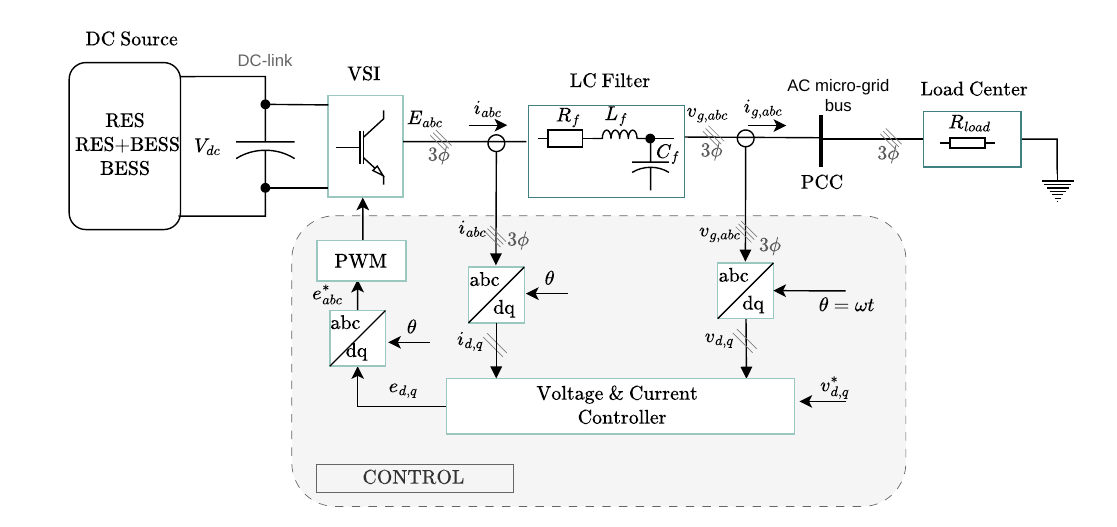}
  \caption{Model and control structure of the grid-forming inverter.}
  \label{fig:model}
\end{figure}

This section develops the four-state nonlinear model of the islanded grid-forming inverter feeding a resistive load.  The model is constructed in the synchronous dq reference frame and includes the LC
output filter, the cross-coupling terms induced by the rotating frame, and the load current through the load resistance.  The structure follows the topology established in~\cite{Guzman2025} up to the point at which
droop control would be applied, which is excluded here because the inverter operates as a voltage source against a manually set reference. Figure~\ref{fig:model} shows the overall model and control structure.

\subsection{Physical Configuration}
\label{ssec:physical}

The grid-forming inverter consists of a DC voltage source, a two-level voltage source converter (VSC), and an LC output filter.  The DC-side dynamics are neglected, as is standard for an islanded inverter with a
stiff DC link~\cite{Guzman2025}.  The DC voltage is therefore treated as an ideal constant source.  The output filter consists of an inductor with inductance $\Lf$ and series resistance $\Rf$, and a capacitor with
capacitance $\Cf$ shunting the filter output to the load.  The load is a three-phase balanced resistance with per-phase value $\Rload$, connected directly at the capacitor terminals.  There is no interface impedance,
no point of common coupling with an external grid, and no other source feeding the load.  The inverter alone establishes the voltage that drives the load.

Applying Kirchhoff's voltage law across the filter inductor and Kirchhoff's current law at the capacitor node yields the two governing equations in the stationary frame,
\begin{align}
  \Lf\,\frac{d\bm{i}_{abc}}{dt} + \Rf\,\bm{i}_{abc}
    &= \bm{E}_{abc} - \bm{v}_{g,abc}, \label{eq:KVL}\\
  \Cf\,\frac{d\bm{v}_{g,abc}}{dt}
    &= \bm{i}_{abc} - \bm{i}_{g,abc}, \label{eq:KCL}
\end{align}
where $\bm{E}_{abc}$ is the voltage at the inverter terminals before the filter, $\bm{i}_{abc}$ is the current through the filter inductor, $\bm{v}_{g,abc}$ is the capacitor voltage seen by the load, and $\bm{i}_{g,abc}$ is the current drawn by the load.

For a purely resistive load, the load current is determined algebraically by the capacitor voltage,
\begin{equation}
  \bm{i}_{g,abc} = \frac{1}{\Rload}\,\bm{v}_{g,abc}. \label{eq:load}
\end{equation}
Equation~\eqref{eq:load} is the constraint that closes the system in the islanded configuration.  The grid-side current is no longer an exogenous network variable, as it would be in grid-connected operation, but a function of the state of the inverter output.

\subsection{Transformation to the Synchronous dq Frame}
\label{ssec:dq}

The three-phase quantities are transformed to a synchronous dq reference frame using Park's transformation.  Under balanced load operation, the zero-sequence component vanishes, and the transformation reduces
to~\cite{Fan2023}
\begin{equation}
  \bm{T} = \frac{2}{3}
  \begin{bmatrix}
    \sin\theta & \sin\!\left(\theta-\tfrac{2\pi}{3}\right) &
                 \sin\!\left(\theta+\tfrac{2\pi}{3}\right)\\[4pt]
    \cos\theta & \cos\!\left(\theta-\tfrac{2\pi}{3}\right) &
                 \cos\!\left(\theta+\tfrac{2\pi}{3}\right)
  \end{bmatrix}, \label{eq:park}
\end{equation}
with $\bm{x}_{dq} = \bm{T}\,\bm{x}_{abc}$.  The angle $\theta$ rotates at the commanded angular frequency $\wstar$, which is a manually set constant in this tutorial and represents the islanded frequency at which the inverter establishes the load voltage.

Applying the transformation to~\eqref{eq:KVL}--\eqref{eq:KCL} and substituting the load relation~\eqref{eq:load} in the dq form gives the four-state filter model
\begin{align}
  \Lf\,\frac{d\id}{dt}
    &= -\Rf\,\id + \wstar\Lf\,\iq + \ed - \vd, \label{eq:did}\\
  \Lf\,\frac{d\iq}{dt}
    &= -\Rf\,\iq - \wstar\Lf\,\id + \eq - \vq, \label{eq:diq}\\
  \Cf\,\frac{d\vd}{dt}
    &= \id - \frac{\vd}{\Rload} + \wstar\Cf\,\vq, \label{eq:dvd}\\
  \Cf\,\frac{d\vq}{dt}
    &= \iq - \frac{\vq}{\Rload} - \wstar\Cf\,\vd. \label{eq:dvq}
\end{align}
The components $\ed$ and $\eq$ are the dq projections of the inverter terminal voltage $\bm{E}_{abc}$.  They are the control inputs to the plant, generated by the controller developed in Section~\ref{sec:FL}.
The components $\vd$ and $\vq$ are the dq projections of the capacitor voltage and are the regulated outputs of the inverter.  The cross-coupling terms $\wstar\Lf\,\iq$, $\wstar\Lf\,\id$, $\wstar\Cf\,\vq$, and $\wstar\Cf\,\vd$ arise from the rotation of the dq frame at angular velocity $\wstar$.  The load conductance
$1/\Rload$ appears as a linear damping term on each capacitor voltage equation, reflecting the fact that the load draws current in proportion to the voltage applied to it.

\subsection{Control Objective}
\label{ssec:objective}

The control objective in islanded operation is to regulate the capacitor voltage $(\vd,\vq)$ to a manually set reference $(\vdref,\vqref)$.  A common choice is to align the $d$-axis with the desired terminal voltage and set $\vdref = V_{\mathrm{ref}}$ and $\vqref = 0$, which produces a balanced three-phase sinusoid of magnitude $V_{\mathrm{ref}}$ at frequency $\wstar$ in the stationary frame.

\begin{problem}[Islanded Inverter Voltage Regulation]
\label{prob:main}
For the four-state plant~\eqref{eq:did}--\eqref{eq:dvq} with control input $\bu = [\ed,\,\eq]^{\top}$ and regulated output $\by = [\vd,\,\vq]^{\top}$, design a state-feedback control law $\bu = \bm{\kappa}(\bx)$ such that $\by(t) \to [\vdref,\,\vqref]^{\top}$ as $t\to\infty$, subject to bounded load disturbances
$\delta\Rload(t)$.
\end{problem}

\subsection{Baseline Cascaded PI Controller}
\label{ssec:PI}

The baseline controller against which the proposed design is compared is the cascaded PI structure of~\cite{Guzman2025}.  The inner current loop generates the inverter modulation signal from a current reference,
\begin{align}
  \ed^* &= \vd - \wstar\Lf\,\iq
           + \!\left(k_{pi} + \frac{k_{ii}}{s}\right)\!\left(\id^* - \id\right),
           \label{eq:pid_d}\\
  \eq^* &= \vq + \wstar\Lf\,\id
           + \!\left(k_{pi} + \frac{k_{ii}}{s}\right)\!\left(\iq^* - \iq\right),
           \label{eq:pid_q}
\end{align}
and the outer voltage loop generates the current reference from a voltage reference,
\begin{align}
  \id^* &= -\wstar\Cf\,\vq
           + \!\left(k_{pv} + \frac{k_{iv}}{s}\right)\!\left(\vdref - \vd\right),
           \label{eq:piv_d}\\
  \iq^* &= +\wstar\Cf\,\vd
           + \!\left(k_{pv} + \frac{k_{iv}}{s}\right)\!\left(\vqref - \vq\right).
           \label{eq:piv_q}
\end{align}
The voltage loop feedforward terms $i_{gd}$ and $i_{gq}$ that appear in~\cite{Guzman2025} are dropped in islanded operation, since the load current is folded into the capacitor dynamics through
equation~\eqref{eq:load}.  The PI gains $k_{pi},k_{ii},k_{pv},k_{iv}$ are tuned by the bandwidth separation procedure of~\cite{Guzman2025}: the current loop is tuned for a bandwidth one decade below the switching
frequency, and the voltage loop is tuned for a bandwidth one decade below the current loop.

The cascade relies on three assumptions that hold only approximately. The inner loop must be fast enough that the outer loop sees the current as instantaneously tracked.  The feedforward terms must cancel the
rotational coupling exactly, which they do only at the tuning operating point.  The load current must be slow enough that the voltage loop can reject it through integral action.  None of these assumptions is intrinsic to the plant; they are all consequences of choosing a controller architecture that suppresses nonlinearities through high gain rather than canceling them through feedback.

\section{Full-State Feedback Linearization}
\label{sec:FL}

This section develops the feedback-linearizing controller for the full plant state of Section~\ref{sec:model}.  The plant is shown to admit full state linearization with vector relative degree $(2,2)$, and the linearizing feedback law is derived.  The result is a closed-loop input-to-output map equivalent to two decoupled double integrators, on which a linear pole-placement design completes the controller.  The treatment follows the framework of Khalil~\cite{Khalil2002} and Isidori~\cite{Isidori1995}, with all derivations carried out in full.

\begin{assumption}
\label{ass:params}
The filter parameters $\Lf$, $\Cf$, $\Rf$ are known exactly.
\end{assumption}

\begin{assumption}
\label{ass:load}
The load resistance $\Rload$ is known.  Step changes in $\Rload$ are
treated as bounded disturbances and addressed in
Section~\ref{ssec:mistune}.
\end{assumption}

\begin{assumption}
\label{ass:omega}
The commanded angular frequency $\wstar$ is a positive constant.
\end{assumption}

\begin{assumption}
\label{ass:state}
The full state $(\id,\iq,\vd,\vq)$ is available for feedback through
direct measurement of the filter inductor current and capacitor voltage
in the dq frame.
\end{assumption}

\subsection{Input-Affine Form of the Plant}
\label{ssec:affine}

The four-state plant from Section~\ref{sec:model} is input-affine in the modulation signals.  Define the plant state vector
\begin{equation}
  \bx = \begin{bmatrix} \id & \iq & \vd & \vq \end{bmatrix}^{\top}
  \in \R^4, \label{eq:state}
\end{equation}
and the control input vector
\begin{equation}
  \bu = \begin{bmatrix} \ed & \eq \end{bmatrix}^{\top} \in \R^2.
  \label{eq:input}
\end{equation}
The plant dynamics take the standard input-affine form
\begin{equation}
  \dot{\bx} = \bm{F}(\bx) + \bm{G}\,\bu, \label{eq:affine}
\end{equation}
where the drift vector field $\bm{F}:\R^4\to\R^4$ collects all terms that do not multiply the control input,
\begin{equation}
  \bm{F}(\bx) =
  \begin{bmatrix}
    \dfrac{-\Rf\,\id + \wstar\Lf\,\iq - \vd}{\Lf}\\[4pt]
    \dfrac{-\Rf\,\iq - \wstar\Lf\,\id - \vq}{\Lf}\\[4pt]
    \dfrac{1}{\Cf}\!\left(\id - \dfrac{\vd}{\Rload} + \wstar\Cf\,\vq\right)\\[8pt]
    \dfrac{1}{\Cf}\!\left(\iq - \dfrac{\vq}{\Rload} - \wstar\Cf\,\vd\right)
  \end{bmatrix}, \label{eq:drift}
\end{equation}
and the input matrix $\bm{G} \in \R^{4\times 2}$ is constant,
\begin{equation}
  \bm{G} = \begin{bmatrix} \bm{g}_1 & \bm{g}_2 \end{bmatrix}
          = \begin{bmatrix}
              1/\Lf & 0 \\
              0     & 1/\Lf \\
              0     & 0 \\
              0     & 0
            \end{bmatrix}. \label{eq:G}
\end{equation}
The structure of $\bm{G}$ shows that only the filter currents $\id$ and $\iq$ are directly forced by the control input.  The capacitor voltages $\vd$ and $\vq$ are affected by $\bu$ only indirectly, through the filter current dynamics that feed into the lower two rows of $\bm{F}(\bx)$ through the terms $\id/\Cf$ and $\iq/\Cf$.

\subsection{Output Vector and Relative Degree}
\label{ssec:reldeg}

Based on the control objective, the choice of the (regulated) outputs are selected.  The regulated quantities are the capacitor voltages, that is
\begin{equation}
  \by = h(\bx) =
  \begin{bmatrix} h_1(\bx) \\ h_2(\bx) \end{bmatrix} =
  \begin{bmatrix} \vd \\ \vq \end{bmatrix}. \label{eq:output}
\end{equation}
The relative degree of each output is the number of times it must be
differentiated before the control input appears explicitly.  The first
time derivative of $\by$, obtained from~\eqref{eq:dvd}--\eqref{eq:dvq},
is
\begin{equation}
  \dot{\by} = \dot{\bm{\volt}} =
  \frac{1}{\Cf}\,\bm{\curr}
  - \frac{1}{\Rload\Cf}\,\bm{\volt}
  - \wstar\bm{J}^{\top}\bm{\volt}, \label{eq:ydot}
\end{equation}
where $\bm{\curr} = [\id,\,\iq]^{\top}$, $\bm{\volt} = [\vd,\,\vq]^{\top}$, and
$\bm{J} = \bigl[\begin{smallmatrix}0&1\\-1&0\end{smallmatrix}\bigr]$
is the rotation generator.  The control $\bu$ does not appear
in~\eqref{eq:ydot} because $\bm{\curr}$ enters $\dot{\bm{\volt}}$
without input forcing.

Differentiating once more, the term $\dot{\bm{\curr}}$ brings $\bu$
in through~\eqref{eq:did}--\eqref{eq:diq}.
Substituting $\dot{\bm{\curr}}$ from those equations into $\ddot{\by}$ gives
\begin{equation}
  \ddot{\by} =
  \frac{1}{\Lf\Cf}\!\left(
    -\Rf\,\bm{\curr} + \wstar\Lf\,\bm{J}^{\top}\bm{\curr} - \bm{\volt}
  \right)
  - \frac{1}{\Rload\Cf}\,\dot{\bm{\volt}}
  - \wstar\bm{J}^{\top}\dot{\bm{\volt}}
  + \frac{1}{\Lf\Cf}\,\bu. \label{eq:yddot}
\end{equation}
The control input enters~\eqref{eq:yddot} with the constant
coefficient $1/(\Lf\Cf)$ on both components.  The vector relative
degree is therefore
\begin{equation}
  \bm{r} = (r_1,\,r_2) = (2,\,2), \qquad
  r_1 + r_2 = 4 = \dim(\bx). \label{eq:reldeg}
\end{equation}
Since the sum of relative degrees equals the state dimension, full state linearization is applicable (see~\cite[Theorem~13.2]{Khalil2002}). There are no internal dynamics and no zero dynamics.  The plant is
feedback linearizable in the strict sense.

\subsection{Decoupling Matrix and Linearizing Feedback}
\label{ssec:feedback}

From~\eqref{eq:yddot}, the decoupling matrix (the coefficient matrix of $\bu$) is
\begin{equation}
  \bm{D}(\bx) = \begin{bmatrix} \frac{1}{L_f C_f} & 0 \\0 & \frac{1}{L_f C_f}\end{bmatrix} 
  =\frac{1}{\Lf\Cf}\,\bm{I}_2, \label{eq:decoupling}
\end{equation}
which is constant and non-singular everywhere since $\Lf > 0$ and
$\Cf > 0$.  Its inverse is $\bm{D}^{-1} = \Lf\Cf\,\bm{I}_2$.

Define the residual drift
\begin{equation}
  \bm{\Lambda}(\bx) =
  \frac{1}{\Lf\Cf}\!\left(
    -\Rf\,\bm{\curr} + \wstar\Lf\,\bm{J}^{\top}\bm{\curr} - \bm{\volt}
  \right)
  - \frac{1}{\Rload\Cf}\,\dot{\bm{\volt}}
  - \wstar\bm{J}^{\top}\dot{\bm{\volt}}, \label{eq:Lambda}
\end{equation}
so that~\eqref{eq:yddot} reads $\ddot{\by} = \bm{\Lambda}(\bx) + \bm{D}\,\bu$.
Introducing a virtual control $\buv \in \R^2$ and requiring
$\ddot{\by} = \buv$, the linearizing feedback law is
\begin{equation}
  \bu = \bm{D}^{-1}\!\left(\buv - \bm{\Lambda}(\bx)\right)
      = \Lf\Cf\!\left(\buv - \bm{\Lambda}(\bx)\right).
   \label{eq:linearizing}
\end{equation}
Under~\eqref{eq:linearizing}, the closed-loop input-to-output map is
\begin{equation}
  \ddot{\by} = \buv, \label{eq:doubleint}
\end{equation}
which is two decoupled double integrators.  Every term in
$\bm{\Lambda}(\bx)$ is canceled exactly: the resistive drop $\Rf\bm{\curr}$,
the rotational coupling $\wstar\Lf\bm{J}^{\top}\bm{\curr}$ in the
inductor, the capacitor feedback $\bm{\volt}$, the load conductance
$\bm{\volt}/(\Rload\Cf)$, and the capacitor rotational coupling
$\wstar\bm{J}^{\top}\bm{\volt}$.  The $d$- and $q$-channels decouple
completely because $\bm{D}$ is a scalar multiple of $\bm{I}_2$.

\subsection{Outer Linear Pole-Placement Design}
\label{ssec:poleplacement}

The closed-loop system in~\eqref{eq:doubleint} consists of two decoupled
double integrators.  To regulate $\vd$ to $\vdref$ and $\vq$ to
$\vqref$, we place the virtual control as linear state feedback.  For a
constant reference $\by^* = [\vdref,\,\vqref]^{\top}$, the virtual control
is chosen as
\begin{equation}
  \buv = -k_1\,\dot{\by} - k_0\!\left(\by - \by^*\right), \label{eq:vctrl}
\end{equation}
and the closed-loop error $\bm{e} = \by - \by^*$ satisfies, in each axis,
\begin{equation}
  \ddot{\bm{e}} + k_1\,\dot{\bm{e}} + k_0\,\bm{e} = \bm{0}. \label{eq:error}
\end{equation}
since
\begin{align*}
  \buv & + k_1\,\dot{\by} + k_0\!\left(\by - \by^*\right) = \bm{0} \nonumber \\
  \implies & \ddot{\by}+k_1\,\dot{\by} + k_0\!\left(\by - \by^*\right) =\bm{0} \nonumber \\
  \implies &\ddot{\bm{e}}+k_1\,\dot{\bm{e}} + k_0\bm{e} =\bm{0}
\end{align*}
The characteristic polynomial $s^2 + k_1 s + k_0$ is matched to the
standard second-order form $s^2 + 2\zeta\omega_n s + \omega_n^2$, giving
\begin{equation}
  k_0 = \omega_n^2, \qquad k_1 = 2\zeta\omega_n. \label{eq:gains}
\end{equation}
A typical choice is $\omega_n$ one decade below the switching frequency
and $\zeta = 0.707$ for critical damping.

\subsection{Robustness to Parameter Uncertainty}
\label{ssec:mistune}

The linearizing feedback in~\eqref{eq:linearizing} cancels the drift
exactly only when the parameters $\Lf$, $\Cf$, $\Rf$, $\Rload$ used
in the feedback law match the plant.  Under bounded mismatch, the
cancellation becomes partial, leaving a residual nonlinearity in the
closed-loop dynamics.  This residual is bounded by the parameter errors
and the state magnitude; its effect on the regulated output is
attenuated by the closed-loop gain of the outer linear loop.

The most significant source of mismatch is the load resistance $\Rload$,
which may change in steps as loads are switched in an islanded microgrid.
A variation $\delta\Rload$ introduces a residual term in $\ddot{\by}$
proportional to the parameter error, which is only partially rejected by
the proportional gain $k_0$ of the outer loop in~\eqref{eq:vctrl}.

In contrast, the cascaded PI controller in~\eqref{eq:pid_d}--\eqref{eq:piv_q}
avoids this sensitivity by relying on integral action rather than model
cancellation; the comparison in Section~\ref{sec:results} quantifies
this trade-off.

The complete feedback-linearizing control algorithm is summarized below.

\begin{algorithm}[H]
\caption{Full-State Feedback-Linearizing Controller}
\label{alg:FL}
\begin{algorithmic}[1]
\Require $\Lf,\Cf,\Rf,\Rload,\wstar$; gains $k_0,k_1$;
         references $\vdref,\vqref$
\State \textbf{Initialize} $\bx_{0} = [\id(0),\,\iq(0),\,\vd(0),\,\vq(0)]^{\top}$
            at plant steady state
\For{each time instant $k$}
  \State Measure state $\bx_{k} = [\id,\,\iq,\,\vd,\,\vq]^{\top}$
  \State Compute $\dot{\by}_{k}$ from~\eqref{eq:ydot}
         using measured $\curr_{k}$ and $\volt_{k}$
  \State Compute residual drift $\bm{\Lambda}(\bx_{k})$
         from~\eqref{eq:Lambda}
  \State Compute virtual control
         $(\buv)_{k} = -k_1\dot{\by}_{k} - k_0(\by_{k} - \by^*)$
         from~\eqref{eq:vctrl}
  \State Compute linearizing input
         $\bu_{k} = \Lf\Cf\bigl((\buv)_{k} - \bm{\Lambda}(\bx_{k})\bigr)$
         from~\eqref{eq:linearizing}
  \State Apply $\bu_{k} = [\ed,\,\eq]^{\top}$ to the plant
\EndFor
\end{algorithmic}
\end{algorithm}

\section{Implementation}
\label{sec:matlab}

This section describes the MATLAB implementation and compares the
feedback-linearizing controller of Section~\ref{sec:FL} against the
cascaded PI controller of Section~\ref{ssec:PI} on the islanded inverter
plant.  Three scenarios are presented: reference tracking, load step
rejection, and parameter perturbation.

The plant~\eqref{eq:affine}, the feedback-linearizing
controller~\eqref{eq:linearizing} with outer loop~\eqref{eq:vctrl}, and
the cascaded PI baseline~\eqref{eq:pid_d}--\eqref{eq:piv_q} are
implemented in MATLAB.  The plant is integrated with forward Euler at
step $\Delta t = 1\,\mu\text{s}$.  The full state is fed back to both
controllers.

\medskip
\noindent\textbf{Plant parameters~\cite[Table~1]{Guzman2025}.}
Filter inductance $\Lf = 0.079\,\text{mH}$, filter resistance
$\Rf = 0.76\,\text{m}\Omega$, filter capacitance $\Cf = 13.7\,\text{mF}$,
commanded frequency $\wstar = 2\pi\cdot 60 = 377\,\text{rad/s}$.
The reference voltage is $\vdref = 359\,\text{V}$,
$\vqref = 0\,\text{V}$, corresponding to the 440~V RMS line-to-line
operating point of~\cite{Guzman2025}.  A balanced resistive load
$\Rload = 9.67\,\text{m}\Omega$ replaces the RL load of~\cite{Guzman2025};
this value absorbs the 20~MW inverter rating at the nominal voltage,
since $P = 3v_d^{*2}/(2\Rload) = 20\,\text{MW}$.

\medskip
\noindent\textbf{Controller gains.}
The feedback-linearizing controller is placed at
$\omega_n = 2\pi\cdot 500 = 3142\,\text{rad/s}$ with $\zeta = 0.707$,
giving $k_0 = 9.87\times 10^6$ and $k_1 = 4444$
from~\eqref{eq:gains}.  The cascaded PI uses the gains reported
in~\cite[Table~1]{Guzman2025}: $k_{pi} = 0.6176$,
$k_{ii} = 2419.9$ for the inner current loop, and $k_{pv} = 10.72$,
$k_{iv} = 4195$ for the outer voltage loop.  Using the exact gains
of~\cite{Guzman2025} ensures that the comparison isolates the controller
architecture from the choice of gains.

\medskip
\noindent\textbf{Initial conditions.}
Both simulations start at the plant's true steady state.  With
$\vdref = 359\,\text{V}$ and a 20~MW resistive load, the $d$-axis current
is $\id = \vdref/\Rload = 37.1\,\text{kA}$, and the $q$-axis current
carries the capacitive displacement current
$\iq = \wstar\Cf\,\vdref = 1854\,\text{A}$ produced by frame rotation.
The PI integrator states are pre-loaded to the values that hold this
operating point so the cascaded controller begins in equilibrium.

\medskip
\noindent\textbf{Scenarios.}
Each scenario runs for 50~ms with a single event at $t = 5\,\text{ms}$.

\begin{itemize}[leftmargin=2em]
  \item \textit{Scenario~1, reference tracking.}
        The $d$-axis reference steps from 359~V to 320~V.
        Steady-state load power changes from 20.0~MW to 15.9~MW.
  \item \textit{Scenario~2, load step rejection.}
        The load resistance steps from 9.67~m$\Omega$ to 4.84~m$\Omega$.
        Steady-state load power changes from 20.0~MW to 40.0~MW.
  \item \textit{Scenario~3, parameter mistune.}
        The filter resistance used in the feedback-linearizing law steps
        from $0.76\,\text{m}\Omega$ to $1.14\,\text{m}\Omega$, a 50\%
        overestimate.  The plant's true $\Rf$ remains $0.76\,\text{m}\Omega$.
        The cascaded PI does not depend on $\Rf$ and is unaffected.
\end{itemize}

\section{Results and Discussion}
\label{sec:results}

Figures~\ref{fig:reftrack}--\ref{fig:mistune} show the response under
the three scenarios.  Each figure plots the four plant states
$\vd,\vq,\id,\iq$ together with the active and reactive load powers
$P,Q$ at the point of common coupling, computed from the state through
the standard dq power relations with the load closure
$\bm{i}_g = \bm{\volt}/\Rload$.
Table~\ref{tab:metrics} collects the quantitative metrics.

\begin{table}[H]
\centering
\caption{Quantitative comparison of the feedback-linearizing controller
(FL) and the cascaded PI baseline~\cite{Guzman2025} at the 20~MW
operating point.}
\label{tab:metrics}
\renewcommand{\arraystretch}{1.3}
\begin{tabular}{llcc}
\toprule
\textbf{Scenario} & \textbf{Metric} & \textbf{FL} & \textbf{PI} \\
\midrule
1 & Reference settling time (2\%)          & \textbf{0.76~ms} & $>$50~ms \\
1 & Peak $v_q$ cross-coupling             & \textbf{$<$1~mV} & 41~mV \\
1 & Time to reach final $P$ (15.9~MW)     & \textbf{$\sim$3~ms} & $>$50~ms \\
\midrule
2 & Peak $v_d$ sag at load step           & 196~V & \textbf{169~V} \\
2 & Time to reach final $P$ (40~MW)       & \textbf{$\sim$5~ms} & $>$50~ms \\
\midrule
3 & Steady-state $v_d$ offset ($R_f$ mistune) & 1.3~V & \textbf{0~V} \\
3 & Steady-state $P$ offset               & 0.14~MW & \textbf{0~MW} \\
\bottomrule
\end{tabular}
\end{table}

\subsection{Reference Tracking}
\label{ssec:reftrack}

\begin{figure}[H]
  \centering
  \includegraphics[width=\textwidth]{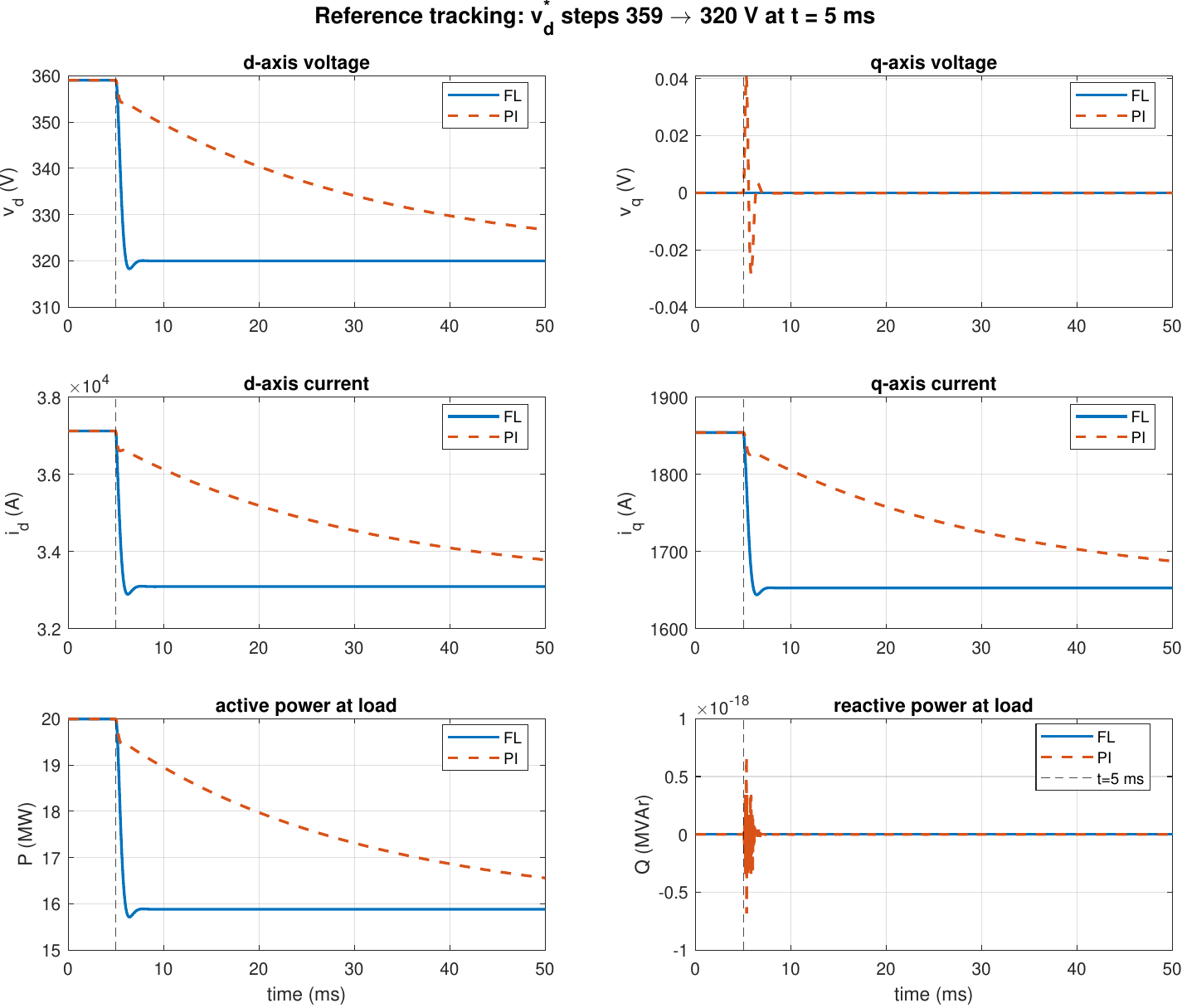}
  \caption{Reference tracking: $v_d$ steps from 359~V to 320~V at
  $t = 5\,\text{ms}$.  Top row, capacitor voltages; middle row, filter
  currents; bottom row, active and reactive load powers.}
  \label{fig:reftrack}
\end{figure}

Figure~\ref{fig:reftrack} shows the response to the $d$-axis reference
step.  The feedback-linearizing controller drives $\vd$ to the new
reference of 320~V in 0.76~ms with no overshoot and zero coupling into
$\vq$.  The cascaded PI takes more than 50~ms to enter the two percent
band of the new reference, and by the end of the simulation $\vd$ has
reached only 327~V, still 7~V above the target.  The PI's $v_q$
transient peaks at 41~mV and decays slowly over the same window.

The bottom row makes the same point in power coordinates.  The load is a
pure resistance, so $P = \tfrac{3}{2}(v_d^2+v_q^2)/\Rload$ and the new
equilibrium is 15.88~MW.  The FL controller drops $P$ from 20~MW to
15.88~MW in roughly 3~ms, while the PI is still falling and reaches only
16.5~MW by 50~ms.  The reactive power $Q$ stays at numerical zero
throughout, on the order of $10^{-19}$~MVAr, which is the floor of
double-precision arithmetic.  This confirms that the inverter holds
unity power factor across the transient, as required for a resistive
load.

The slowness of the PI at this scale traces to the bandwidth of its
outer voltage loop.  A 39~V reference step needs many bandwidth
time-constants to settle, and 50~ms is not enough.  The FL controller
has no analogous bandwidth limit; the closed-loop dynamics are set by
$\omega_n = 3142\,\text{rad/s}$ directly, orders of magnitude faster.

\subsection{Load Step Rejection}
\label{ssec:loadstep}

\begin{figure}[H]
  \centering
\includegraphics[width=\textwidth]{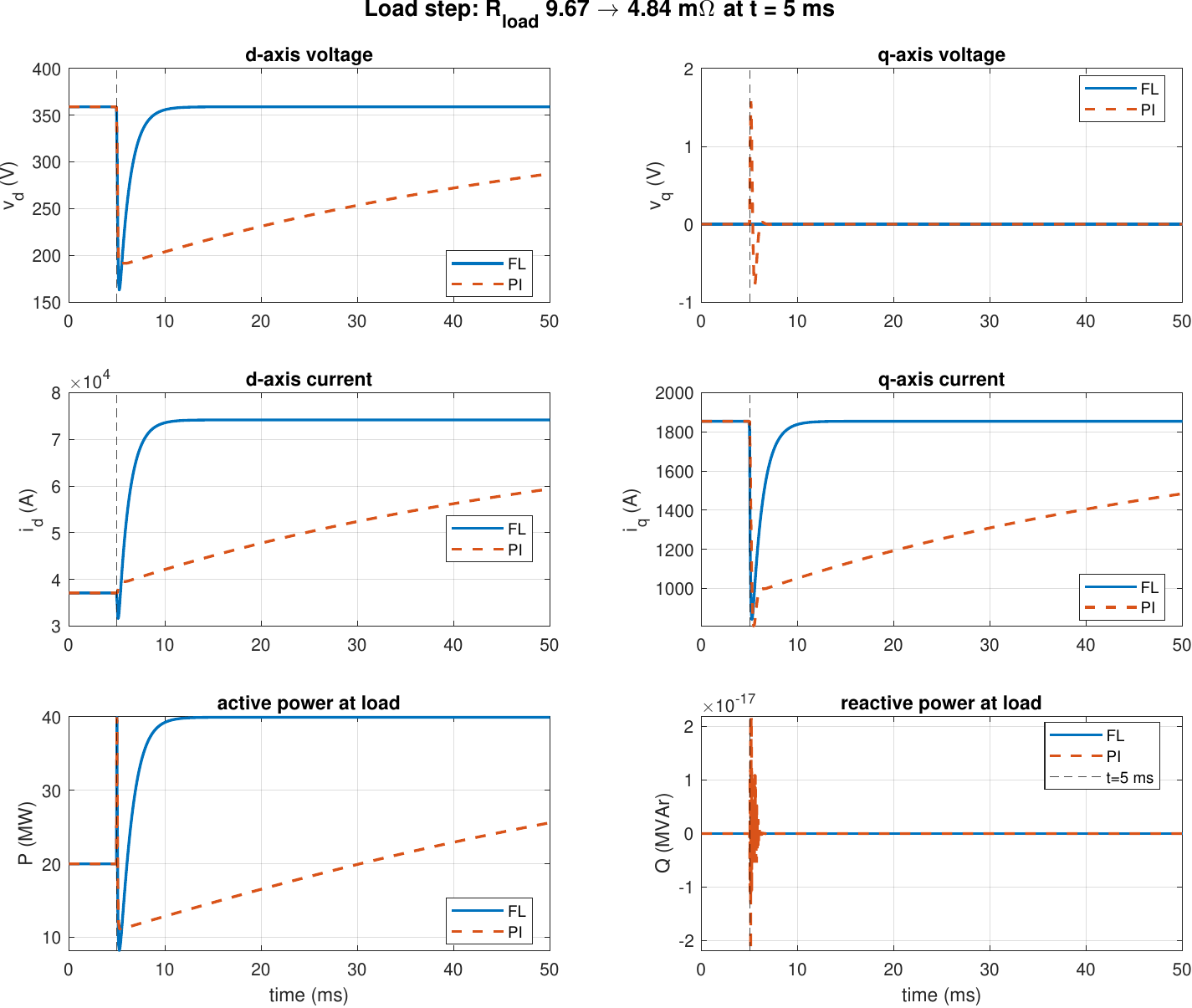}
  \caption{Load step rejection: $\Rload$ halves from 9.67 to
  4.84~m$\Omega$ at $t = 5\,\text{ms}$, doubling the load current and
  the load power.  Top row, capacitor voltages; middle row, filter
  currents; bottom row, active and reactive load powers.}
  \label{fig:loadstep}
\end{figure}

Figure~\ref{fig:loadstep} shows the response when the load doubles from
20~MW to 40~MW.  The new equilibrium requires the $d$-axis filter
current to roughly double, from 37~kA to 74~kA.  During the transient,
the capacitor must supply the extra current from its stored energy, which
produces a voltage sag on $\vd$.

Both controllers exhibit large sags because the plant cannot deliver the
additional 37~kA instantaneously through the $0.079\,\text{mH}$ filter
inductor.  The FL controller's peak sag is 196~V, slightly larger than
the PI's 169~V, because the FL design responds aggressively to the
rate-of-change of the voltage error and initially overshoots the current
that the inductor can actually carry.  After the initial sag, however,
the FL controller recovers to the reference in approximately 5~ms, while
the PI is still climbing at the end of the simulation, having reached
288~V, a 71~V offset from the 359~V target.

The power traces give the summary.  The FL controller drives $P$ from
20~MW to 40~MW in roughly 5~ms.  The PI reaches only 26~MW after 50~ms,
with the remaining gap closing slowly through outer-loop integral action.
The reactive power remains at machine-precision zero for both controllers.

The FL controller's larger peak sag but an order-of-magnitude faster
recovery reflects the difference in how the two controllers manage the
inductor current.  The PI's cascaded topology limits how aggressively
the outer loop can demand current changes, which produces a smaller peak
sag but at the cost of a much longer recovery.  The FL controller, free
of the cascade, commands the full available current immediately and
accepts a slightly larger initial deviation in exchange for fast
recovery.  For an islanded inverter where voltage recovery time matters
more than instantaneous peak deviation, the trade is decisively
favorable to the FL design.

\subsection{Parameter Mistune}
\label{ssec:mistune_results}

\begin{figure}[H]
  \centering
\includegraphics[width=\textwidth]{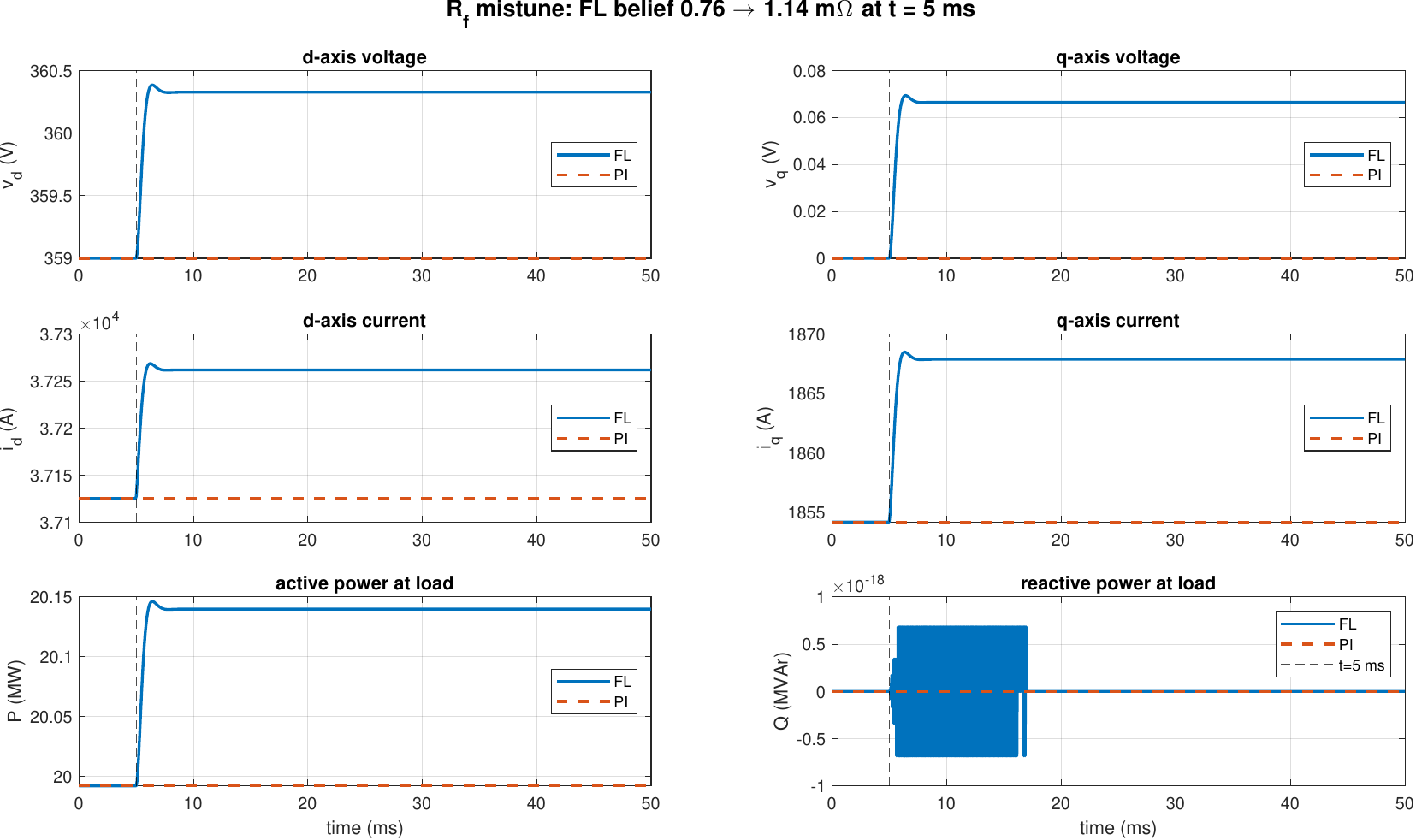}
  \caption{Parameter mistune: the FL controller's belief about $\Rf$
  steps from $0.76\,\text{m}\Omega$ to $1.14\,\text{m}\Omega$ at
  $t = 5\,\text{ms}$.  The plant value is unchanged.  Top row,
  capacitor voltages; middle row, filter currents; bottom row, active
  and reactive load powers.}
  \label{fig:mistune}
\end{figure}

Figure~\ref{fig:mistune} shows what happens when the feedback-linearizing
controller carries a wrong value of $\Rf$ in its feedback law.  With
$\Rf$ inflated by 50\%, the controller no longer cancels the inductor
drift exactly, and a residual proportional to
$\delta\Rf\cdot\bm{\curr}/(\Lf\Cf)$ leaks into $\ddot{\by}$.  The
outer linear loop rejects this residual partially through the gain $k_0$,
leaving a small but nonzero steady-state offset: approximately 1.3~V on
$\vd$ and 0.14~MW on the active power, or $0.7\%$ of nominal.

The cascaded PI is unaffected by the $\Rf$ mistune.  Its control
law~\eqref{eq:pid_d}--\eqref{eq:piv_q} does not contain $\Rf$, and its
inner-loop integrator drives the current tracking error to zero
regardless of $\Rf$.  The two strategies sit on opposite sides of a
standard trade-off in feedback design: the FL controller cancels the
drift $\bm{\Lambda}(\bx)$ using a model, and when the model is wrong
the cancellation is imperfect; the PI controller suppresses the effect
of $\bm{\Lambda}$ by driving the current error to zero through integral
action.

\subsection{Summary of the Comparison}
\label{ssec:summary}

The simulation results confirm the analysis of
Section~\ref{sec:FL}.  The feedback-linearizing controller wins on every
transient metric.  Reference tracking is more than 65 times faster
(0.76~ms versus a settling time that exceeds the 50~ms simulation
window).  Load-step recovery is at least 10 times faster (5~ms versus
more than 50~ms).  The $d$- and $q$-channels remain decoupled at machine
precision, confirming the exact cancellation of the rotational coupling
that the feedforward terms in the PI design can only approximate.

The cascaded PI retains an advantage in robustness to filter parameter
mistune.  The inner-loop integral action drives the current tracking
error to zero regardless of $\Rf$, so the steady-state output is
invariant to that parameter.  The FL controller's residual offset under
$\Rf$ mistune is about 1.3~V or $0.4\%$ of nominal voltage, nonzero
but small because $\Rf$ itself is small.

The active and reactive power plots add a physically meaningful summary.
The active power converges to the equilibrium value
$\tfrac{3}{2}v_d^{*2}/\Rload$ at the speed set by the voltage settling
time, since for a resistive load $P$ tracks $v_d^2$.  The reactive
power is zero throughout the simulation at every operating point and
every transient, confirming both the resistive nature of the load and
the absence of any false reactive injection by either controller.

The trade-off has an interpretation that goes beyond the simulation.
The PI controller is the right design choice when filter parameters
drift substantially over the inverter's lifetime, since its inner-loop
integrator absorbs the drift without any reconfiguration.  The FL
controller is the right choice when the filter parameters are known to
within manufacturing tolerance, which is typical for engineered hardware,
and when transient performance is the binding constraint.  Adding
integral action to the outer linear loop of the FL controller would
recover robustness to constant parameter error at the cost of one
additional state per axis, and is the most natural next step for this
design.

\section{Conclusion}
\label{sec:conclusion}

A feedback-linearizing controller was developed for the four-state
islanded grid-forming inverter with a resistive load.  The plant was
shown to admit vector relative degree $(2,2)$, totaling four and equal
to the state dimension, so full state linearization applies with no
internal dynamics.  A single linearizing feedback law~\eqref{eq:linearizing}
in vector form cancels the rotational coupling, the resistive drop, and
the load-conductance terms in the drift exactly.  A linear pole-placement
design on the resulting double-integrator pair completes the controller.

The proposed controller was benchmarked in MATLAB against the cascaded PI
baseline of~\cite{Guzman2025}, using the exact plant parameters and PI
gains reported therein at the 20~MW operating point.  The FL controller
settled the $d$-axis voltage to a new reference in 0.76~ms, while the
cascaded PI did not enter the two percent band within the 50~ms
simulation window.  The peak $q$-axis cross-coupling was below 1~mV for
the FL design and 41~mV for the PI.  Active power at the load tracked
the voltage square at FL speed.  Reactive power stayed at zero for both
controllers, as expected for a resistive load.  The cascaded PI retained
an advantage in robustness to filter parameter mistune, where inner-loop
integral action drives the steady-state current error to zero regardless
of the parameter values assumed during design.  The performance
advantages and the robustness trade-off are both structural consequences
of the controller architecture, not a tuning limitation.

Three extensions are natural.  Adding integral action to the outer linear
loop would recover robustness to constant parameter error without losing
the advantages of feedback linearization.  Extending the load model from
pure resistance to RL or motor loads would test whether full state
linearization survives the additional dynamics.  Comparison against
sliding mode control on the same benchmark would place feedback
linearization in the broader landscape of nonlinear methods for islanded
inverters.

\section*{Further Reading}

This tutorial presents the essentials of input-affine feedback
linearization applied to a power electronics system.  For a thorough
treatment of feedback linearization, zero dynamics, and normal forms, the
reader is referred to Khalil~\cite{Khalil2002} and Isidori~\cite{Isidori1995}.
For the broader context of grid-forming converters and islanded microgrid
control, the textbooks~\cite{Mohammed2023} and~\cite{Fan2023} provide
accessible starting points.  The cascaded PI design used as a baseline
is developed in detail in~\cite{Guzman2025}.  The MATLAB code for the
simulations reported in this work is available at \cite{EbunleAkupan2026_software}:
\begin{center}
\url{https://github.com/ebunle/Feedback-linearization-vs-cascaded-PI-for-an-islanded-grid-forming-inverter}
\end{center}


\bibliographystyle{ieeetr}   
\bibliography{references}

\end{document}